\documentclass[a4paper]{jpconf}
\usepackage[latin9]{inputenc}
\usepackage{amsbsy}
\usepackage{amssymb}
\usepackage{graphicx}
\usepackage[unicode=true,pdfusetitle,
 bookmarks=true,bookmarksnumbered=false,bookmarksopen=false,
 breaklinks=false,pdfborder={0 0 1},backref=false,colorlinks=false]
 {hyperref}

\makeatletter

\pdfpageheight\paperheight
\pdfpagewidth\paperwidth

\providecommand{\tabularnewline}{\\}

\usepackage{jour-abbrev}
\def\maketitle{}

\makeatother

\begin{document}
\global\long\def\Mbh{M_{\bullet}}
\global\long\def\Ms{M_{\star}}
\global\long\def\Rs{R_{\star}}
\global\long\def\Mo{M_{\odot}}
\global\long\def\Ns{N_{\star}}
\global\long\def\ns{n_{\star}}
\global\long\def\SgrA{\mathrm{Sgr\,}A^{\star}}

\title{EMRIs and the relativistic loss-cone:\\
The curious case of the fortunate coincidence}

\author{Tal Alexander}

\maketitle
\address{Dept. of Particle Physics and Astrophysics, Weizmann Institute of Science, Rehovot, Israel}

\ead{tal.alexander@weizmann.ac.il}
\begin{abstract}
Extreme mass ratio inspiral (EMRI) events are vulnerable to perturbations
by the stellar background, which can abort them prematurely by deflecting
EMRI orbits to plunging ones that fall directly into the massive black
hole (MBH), or to less eccentric ones that no longer interact strongly
with the MBH. A coincidental hierarchy between the collective resonant
Newtonian torques due to the stellar background, and the relative
magnitudes of the leading-order post-Newtonian precessional and radiative
terms of the general relativistic 2-body problem, allows EMRIs to
decouple from the background and produce semi-periodic gravitational
wave signals. I review the recent theoretical developments \cite{bar+14}
that confirm this conjectured fortunate coincidence \cite{hop+06a},
and briefly discuss the implications for EMRI rates, and show how
these dynamical effects can be probed locally by stars near the Galactic
MBH.
\end{abstract}

\section{Introduction}

\label{s:introduction}Extreme mass ratio inspiral (EMRI) gravitational
wave (GW) emission events, where a stellar mass black hole (BH) of
mass $\Ms$ gradually inspirals into a massive BH (MBH) of mass $\Mbh$,
are one of the main classes of anticipated extra-galactic low-frequency
GW sources. Since $Q=\Mbh/\Ms\gg1$, such BHs probe spacetime near
the MBH almost as test particles, and thereby offer an opportunity
to test general relativity (GR) under theoretically favorable conditions.

However, the inspiral process is not simply a 2-body problem in the
strong gravity regime. The MBH is surrounded by ${\cal O}(Q)$ background
stars inside its radius of dynamical influence $r_{h}\sim G\Mbh/\sigma^{2}$,
where $\sigma^{2}$ is the stellar velocity dispersion in the host
galaxy's spheroidal. The complex dynamics of the stellar cluster around
the MBH, that supplies stars that fall into the MBH, whether directly
on plunging orbits, or indirectly on inspiraling ones, also interferes
with the idealized 2-body motion of the light BH relative to the MBH,
and can even suppress inspiral altogether. 

I show here that EMRIs are in fact possible by virtue of a three-way
fortunate coincidence between the magnitude and timescale of collective
Newtonian effects due to the background stars, which give rise to
strong resonant gravitational torques, and those of the leading-order
precessional and radiative post-Newtonian (PN) terms of GR \cite{hop+06a}.
This curious coincidence raises several questions: How fine-tuned
is it? Is this a specific feature of Einstein's GR, or is it generic
to a larger class of theories of strong gravity? Do still-viable alternatives
to GR predict substantially different plunge / inspiral branching
ratios? Could the very detection of an EMRI in the future rule out
some of them? These questions remain open at this time.

Another consequence of the fortunate coincidence is that rapid relaxation
by resonant torques in the symmetric potential near a MBH \cite{rau+96},
which can dominate slow 2-body relaxation in some regions of phase
space \cite{bar+16}, is ultimately not very important in determining
the steady-state loss-rates (e.g. by tidal disruption or GW inspiral).
By a coincidence that can only be understood in the context of the
relativistic loss-cone dynamics, early naive estimates that included
only a partial treatment of Newtonian dynamics (2-body relaxation,
but not resonant torques) and of GR dynamics (GW dissipation but not
in-plane Schwarzschild precession), yielded correct order of magnitude
loss-rates estimates \cite{ale17}.

Here I briefly review the recent developments in the understanding
and modeling of the relativistic loss-cone that lead to these insights,
focusing on the relevance for EMRI rates, and on the prospects of
probing these dynamical mechanisms locally, near the Milky Way's MBH
$\SgrA$.

\subsection{The loss-cone problem}

\label{ss:losscone}

Orbits whose periapse lies inside the tidal radius $r_{t}=\Rs Q^{1/3}$
(where $\Rs$ is the stellar radius) or inside the last stable orbit
$r_{\bullet}\simeq8r_{g}$ (where $r_{g}=G\Mbh/c^{2})$ are classified
as plunges. In terms of the normalized angular momentum $j=J/\sqrt{G\Mbh a}=1-e^{2}$
(where $a$ is the orbital semimajor axis (sma) and $e$ the eccentricity),
the corresponding plunge conditions are $j<j_{lc}=j_{t}\simeq\sqrt{2r_{t}/a}$
or $j<j_{lc}=j_{\bullet}=\sqrt{16r_{g}/a}$. Since the set of velocity
vectors that take a star from position $\boldsymbol{r}$ to the MBH
span a cone, the set of such orbits are called the loss-cone. Stars
in the loss-cone are destroyed in less than an orbital period. In
steady state, new stars have to be deflected to loss-cone orbits,
and therefore the loss-cone problem is the problem of how a stellar
system around a MBH randomizes. In a spherical potential, where the
angular momentum of individual orbits is conserved in collisionless
motion, randomization is achieved by dynamical relaxation.

\subsection{Loss-cone replenishment by dynamical relaxation}

\label{ss:relaxation}

The classical treatments of the loss-cone problem considered only
slow 2-body relaxation (denoted here also non-resonant relaxation,
NR), which is inherent to any system composed of discrete interacting
particles \cite{fra+76,bah+76,sha+78,bah+77}. A relaxed stellar system
around a MBH settles into a high density powerlaw stellar cusp, $\ns\propto r^{-\alpha}$.
For a single mass population, which we consider here for simplicity,
$\alpha=7/4$ \cite{bah+76}. Near a MBH, the 2-body relaxation timescale
can be expressed as $T_{NR}(a)\sim Q^{2}P(a)/\Ns(a)\log Q$, where
$P$ is the orbital period, $\Ns(a)\sim Q(a/r_{h})^{3-\alpha}$ is
the number of stars with sma $<a$, and $\log Q$ is the Coulomb factor
\cite{bar+14}.

Generally, the 2-body relaxation timescale for changing the angular
momentum $J$ by order of itself, $T_{J}$, is related to the relaxation
timescale for changing the energy $E$ by order of itself, $T_{E}$,
by $T_{J}\sim j^{2}T_{E}$ \cite{ale17}. It is therefore faster to
reach the MBH by relaxation of angular momentum than by relaxation
of energy (equivalently, relaxation of the sma, since $E=G\Mbh/2a$,
using the stellar dynamical convention $E>0$ for bound orbits). 

\begin{figure}
\begin{centering}
\includegraphics[width=0.75\columnwidth]{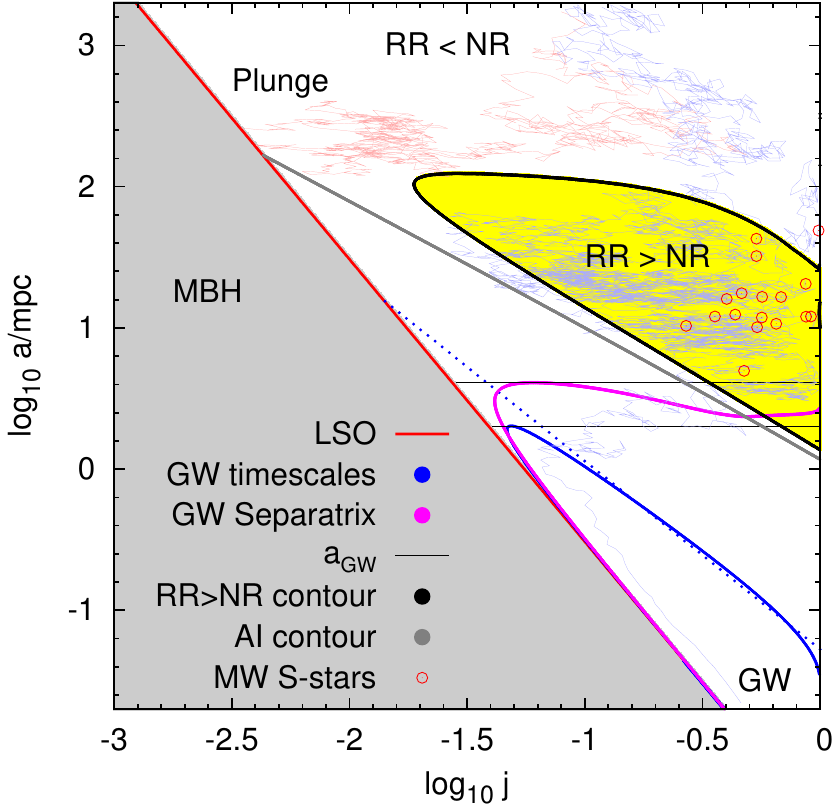}
\par\end{centering}
\caption{\label{f:EJ}A schematic of the relativistic loss cone in $(\log j,\log a)$
phase space for a simplified model of the Milky Way (MW) nucleus with
a $\protect\Mbh=4\times10^{6}\,\protect\Mo$ MBH surrounded by a relaxed
cluster of $10\,\protect\Mo$ stellar BHs \cite{bar+16}. The gray
region to the left, below the last stable orbit (LSO) line demarcates
the region of unstable orbits. Stars diffuse from the Galaxy beyond
the radius of influence at the top ($a\gtrsim2\,\mathrm{pc}$) and
either wander back to the Galaxy, plunge directly into the MBH across
the LSO (the light red track is a Monte Carlo generated track of a
plunge event), or diffuse deeper into the cusp until the cross the
GW separatrix (magenta curve) , where the streamlines of the probability
density flux turn over to the region where GW dissipation is faster
than 2-body relaxation (blue curve). The critical sma $a_{GW}$ (black
horizontal lines, at two possible extreme values) approximately separates
phase space into a region $a>a_{GW}$ where the stellar BHs plunge
directly into the MBH, and a small phase space volume at $a<a_{GW}$
where the BHs inspiral into the MBH. RR is quenched by fast precession
below the AI line (gray line), and the trajectories appear to ``bounce''
against it due to the strong gradient in the effective RR diffusion
coefficient (Fig. \ref{f:AISB}). RR is faster than NR only in a restricted
region of phase space (yellow region), where motion along the $j$-direction
is much faster than in the $a$-direction. Since the RR-dominated
region is well-separated from the plunge and EMRI loss-lines, RR has
a only a small effect on the steady state loss-rates (cf Fig. \ref{f:rates}).
However, it has a strong effect on the orbits of the observed S-stars,
which happen to lie in this phase space region (Sec. \ref{ss:Sstars}).
}
\end{figure}

In the presence of orbital dissipation (e.g. by the emission of GWs,
by tidal deformations of the star close to the MBH, or by hydrodynamical
interactions with a massive accretion disk), the star can also fall
into the MBH by gradually loosing orbital energy. Such orbits are
classified as inspirals. Because inspiral, unlike a plunge, requires
many consecutive periapse passages, such orbits are much more susceptible
to perturbations by the stellar background, unless the inspiral already
starts from a tight orbit around the MBH. phase space is therefore
separated into two regimes (Fig. \ref{f:EJ}). Close to the MBH, below
some critical sma $a_{GW}$ (here we consider only dissipation by
GWs), stars are statistically much more likely to reach the MBH by
inspiral, whereas above $a_{GW}$ and up to $a\sim r_{h}$, plunges
are much more likely. The transition is sharp \cite{hop+05}. The
plunge rate is $R_{p}\sim\Ns(r_{h})/\log(1/j_{lc})T_{E}(r_{h})$,
while the inspiral rate is $R_{i}\sim\Ns(a_{GW})/\log(1/j_{lc})T_{E}(a_{GW})$
\cite{lig+77}. Since the relaxation timescale is typically not a
strong function of distance, the branching ratio $R_{i}/R_{p}\sim\Ns(a_{GW})/\Ns(r_{h})$
reflects the relative number of sources in the small phase space volume
interior to $a_{GW}$ and the much larger one inside $r_{h}$. For
that reason inspiral events are generally much rarer than plunges,
$R_{i}\sim{\cal O}(0.01)R_{p}$ \cite{ale+03b}.

Resonant relaxation \cite{rau+96,hop+06a} is a form of rapid relaxation
of angular momentum that occurs in potentials with a high degree of
symmetry, which restrict orbital evolution (e.g. fixed Keplerian ellipses
in the Newtonian potential of a point mass; fixed orbital planes in
a spherical potential). The approximately spherical, point mass-dominated
potential around a MBH is such a potential. Far enough from the MBH
where GR effects are small, but close enough to it, where deviations
from Keplerian motion due to the distributed stellar mass are also
small, the orbits are nearly-Keplerian ellipses (for $\SgrA$ this
distance scale corresponds to $0.01-0.1$ pc \cite{hop+06a}), which
persist over a long coherence time $T_{c}\gg P$. On that timescale,
the $\Ns$ background stars can be viewed as fixed elliptical mass
``wires''. A test star orbiting the MBH with sma $a$ in the potential
generated by the background, will conserve its energy (since the wires
are stationary), but will be subject to a non-zero residual force
of order $F_{N}\sim{\cal O}(\sqrt{\Ns(a)}G\Ms/a^{2})$, which translates
to a coherent torque $\tau_{N}\sim{\cal O}(\sqrt{\Ns(a)}G\Ms/a)$.
This torque persists in magnitude and direction over time $T_{c}$,
until the small deviations from perfect symmetry accumulate and randomize
the background. The change in angular momentum over the coherence
time, $(\Delta J)_{c}=\tau_{N}T_{c}$ then becomes the mean free path
in $J$-space for a random walk on timescale $t\gg T_{c}$: $\Delta J=(\Delta J)_{c}\sqrt{t/T_{c}}\to\Delta j=\sqrt{t/T_{RR}}$,
where the RR relaxation timescale is $T_{RR}\sim[Q^{2}/\Ns(a)]P^{2}(a)/T_{c}(a)$.

The ratio between the RR and NR relaxation times, $T_{RR}/T_{NR}\sim\log Q(P/T_{c})$
reflects the fact that NR occurs by point-point interactions, and
is boosted by the closest strong interactions, whereas RR occurs by
orbit-orbit interactions, which as extended objects cannot approach
each other arbitrarily close, but are boosted by the long coherence
time. The relevance of RR to the loss-cone problem is due to the fact
that near a MBH, it is possible to have $T_{RR}/T_{NR}\ll1$. This
implies very rapid angular momentum evolution, and specifically $j\to0$,
which allows strong interactions with the MBH. 

\subsection{The fortunate coincidence}

\label{ss:FC}

\begin{figure}

\begin{centering}
\begin{tabular}{cc}
\includegraphics[width=0.5\columnwidth]{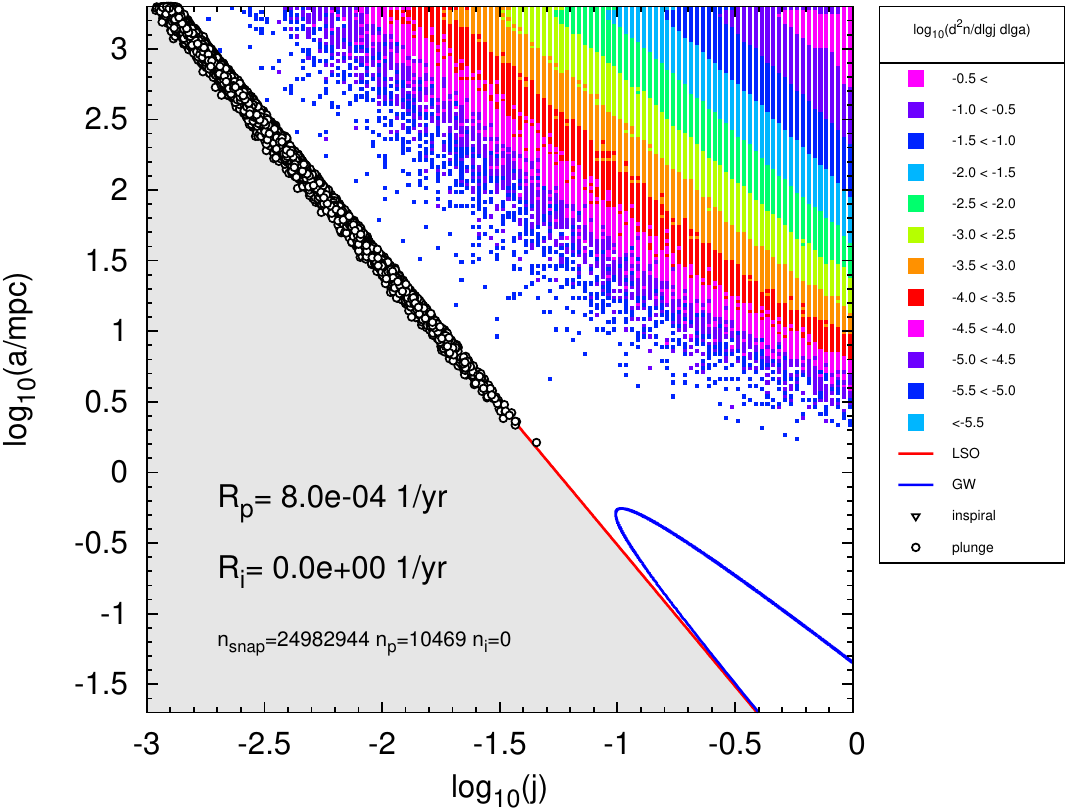} & \includegraphics[width=0.5\columnwidth]{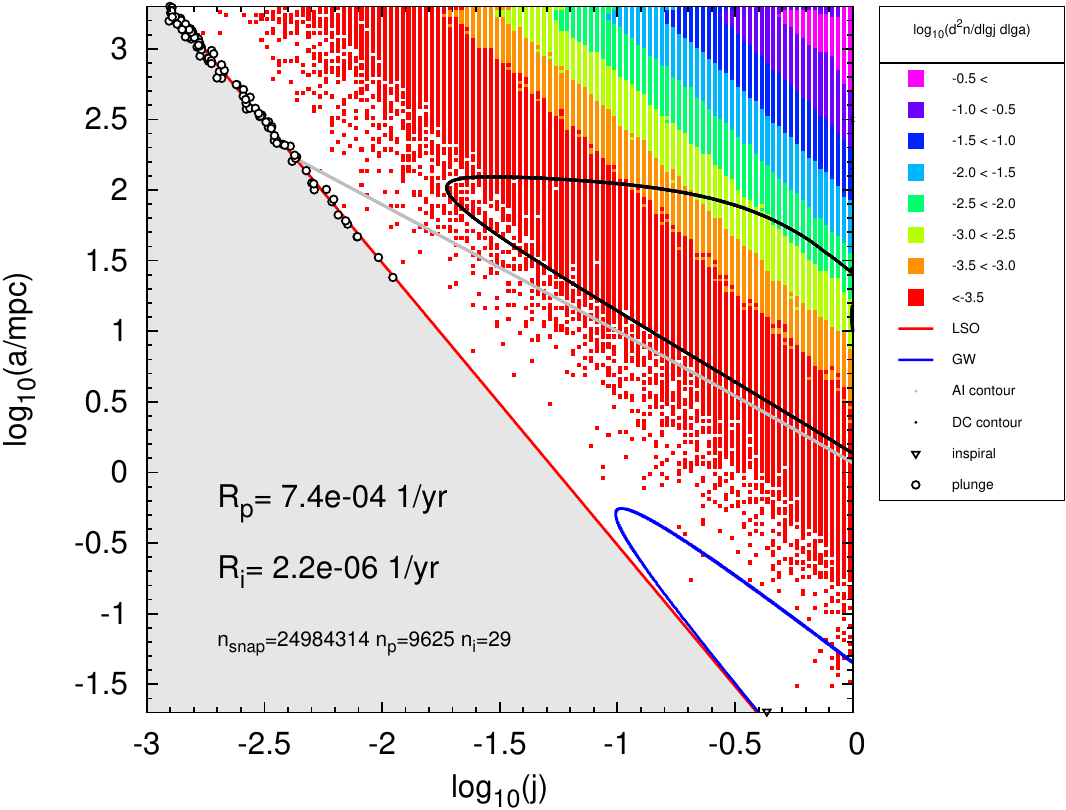}\tabularnewline
\end{tabular}
\par\end{centering}
\caption{\label{f:FC}The steady state density in phase space as derived from
Monte Carlo simulations of loss-cone dynamics for a simplified model
of the Milky Way nucleus surrounded by stellar BHs (see Fig. \ref{f:EJ}),
taking into account the dynamical effects of NR, RR (with the $\eta$-formalism)
and GR (leading order precession and GW) \cite{bar+16}. The plunge
and inspiral rates ($R_{p}$, $R_{i}$) are quoted on the plot. The
end points of a sample of the trajectories are marked by small black
circles. Top: GR Schwarzschild precession turned off. In the absence
of quenching, the strong RR torques sweep all BHs to plunge orbits
well above the GW loss-line, and the EMRI rate drops to zero. Bottom:
All dynamical effects are included. The system reaches a steady state
that is very close to thermal, with a finite inspiral rate $R_{i}\sim{\cal O}(10^{-6}\,\mathrm{yr}^{-1})<0.01R_{p}$.}
\end{figure}

The strong RR torques inside $a_{GW}\ll a\ll r_{h}$, if unquenched,
would drive all stars directly into the MBH on plunge orbits and the
EMRI rate would drop to zero. Hopman \& Alexander conjectured in 2006
\cite{hop+06a} that EMRIs will in fact occur because the PN1 ${\cal O}(\beta^{2}j^{-2})$
GR precession becomes significant enough to quench RR before the PN2.5
${\cal O}(\beta^{5}j^{-7}Q^{-1})$ GW dissipation rate becomes fast
(here $\beta=v/c$). This was later confirmed in $N$-body simulations
\cite{mer+11,bre+14} (but see Sec. \ref{ss:SB}), and was also demonstrated
by Monte Carlo (MC) simulations of diffusion in the relativistic loss-cone
\cite{bar+16} (Sec. \ref{s:etaformalism}). Fig. (\ref{f:FC}) shows
that GR Schwarzschild precession has a drastic effect on the steady
state phase space densities and rates. EMRIs exist because of a fortunate
coincidence in the ranking and phase space dependence of the magnitude
of the RR torques, which are a collective Newtonian effect, and the
magnitudes of GR's leading order precessional and radiative terms,
which are post-Newtonian 2-body effects.

\subsection{An unexpected result}

\label{ss:SB}

The first direct $N$-body simulation with self-consistent post-Newtonian
terms (up to PN2.5, the lowest-order radiative term) \cite{mer+11}
yielded an unanticipated result. Not only did GR precession quench
RR on eccentric orbits, as conjectured, but in fact it appeared that
a barrier in phase space, the ``Schwarzschild Barrier'' (SB), prevented
stars from reaching either the LSO or the GW line (cf Fig. \ref{f:AISB}).
The phase space trajectories of stars seemed to linger near the SB
for about a coherence time, while their orbital parameters oscillated
at at the Schwarzschild precession frequency, before being reflected
back to less eccentric orbits. Larger-scale $N$-body simulations
subsequently confirmed the quenching of RR by GR precession near the
SB \cite{bre+14}. Although an early analysis indicated that the SB
is related to precession under the influence of a randomly changing
uniform force \cite{ale10,mer+11}, the nature and implications of
the SB phenomenon remained controversial. 

\section{The $\boldsymbol{\eta}$-formalism for relativistic loss-cone dynamics}

\label{s:etaformalism}

\begin{figure}
\noindent \centering{}\includegraphics[width=1\columnwidth]{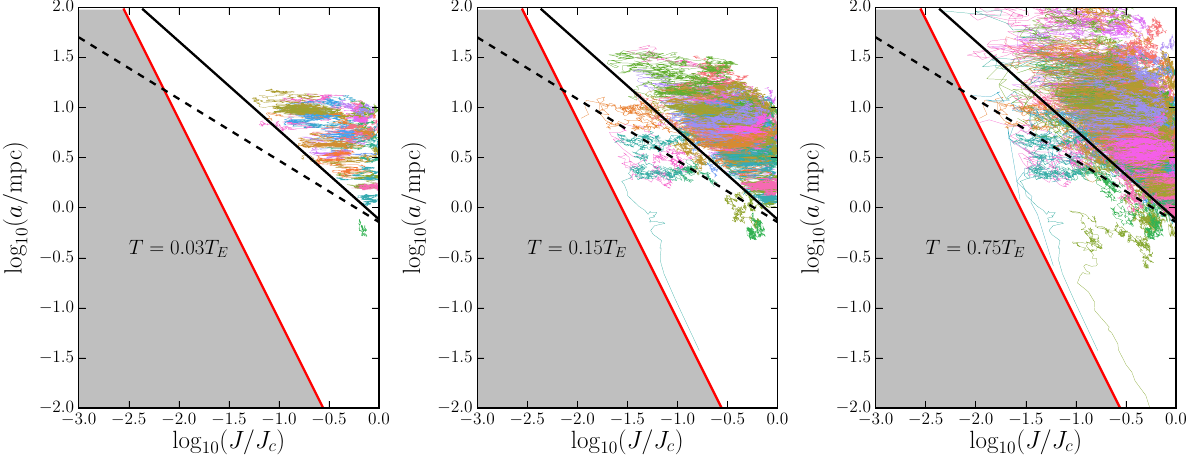}\caption{\label{f:AISB}The nature of the SB / AI line. A time sequence (left
to right) of three snapshots showing the phase space trajectories
of 50 stars of mass $50\,\protect\Mo$ each, in a relaxed cusp ($\alpha=7/4$)
around a MBH of $10^{6}\,\protect\Mo$ \cite{bar+16}. The AI locus
is denoted by a solid black line (the dashed line is the mis-identified
locus of the SB \cite{mer+11}). At times much shorter than the NR
energy relaxation timescale $T_{E}$, evolution by rapid RR dominates,
but can occur only above the AI line. As time progresses, NR, which
is unaffected by AI, populates phase space beyond the AI line, and
ultimately establishes the maximal entropy (thermal) equilibrium on
timescales $\gtrsim{\cal {\cal O}}(T_{E})$, irrespective of AI.}
\end{figure}

Relativistic loss-cone dynamics, and in particular the interplay between
secular precession, the coherent RR torques, and uncorrelated NR,
lie in the difficult-to-treat interface between deterministic Hamiltonian
dynamics and stochastic kinetic theory. The $\boldsymbol{\eta}$-formalism
\cite{bar+14} provides a formal framework for treating this regime
by stochastic equations of motion (EOMs) that describe the dynamics,
by identifying the relevant features of the stochastic effects of
the stellar background on the test star, and importantly, by enabling
the description of RR dynamics by effective diffusion coefficients,
in spite of the fact that the long RR coherence time makes it a manifestly
non-Markovian process that cannot be described, as is, by diffusion. 

The key idea of the $\boldsymbol{\eta}$-formalism is to describe
the effect of the background stars in terms of a time-correlated noise
model. A perturbative expansion of the phase-averaged post-Newtonian
Hamiltonian to leading order shows that the symmetries of the noise
are those of a vector in angular momentum space, $\boldsymbol{\eta}(t)$\footnote{The noise can be approximated as independent of $j$ because the mean
free path in $j$ due to RR is small.}. The dynamics of the system are then primarily determined by the
temporal smoothness (differentiability) of the noise, and by its coherence
time $T_{c}$. Neither of these properties are known at present from
first principles or from $N$-body simulations. However, since the
collective effect of the background is due to the superposition of
many smooth orbital motions, it is likely that the noise is smooth
(infinitely differentiable). A smooth noise must have an maximal frequency
$\nu_{\max}\propto1/T_{c}$, beyond which its power decays rapidly. 

When the GR precession frequency, $\nu_{GR}(a,j)=3(c/r_{g})(r_{g}/a)^{5/2}j^{-2}$
(where $r_{g}=G\Mbh/c^{2}$) is higher than $1/T_{c}$ (i.e. when
the precession period is much shorter than the coherence time), the
residual RR force is effectively constant over a precession period,
so the RR torque on the precessing eccentric star is reversed every
half cycle. Therefore, the net change in $j$ per period is canceled
to high precision\textemdash the rapidly precessing eccentric orbit
is decoupled from the slowly varying stellar background by adiabatic
invariance (AI). This happens along the phase space locus

\begin{equation}
j_{AI}(a)=\sqrt{T_{c}(a)\nu_{GR}(a,j=1)/2\pi}\,.\label{e:jAI}
\end{equation}

The stochastic EOMs allow to evolve in time the phase space trajectory
of a test star, for a given random realization of the background noise.
The fact that the noise is approximately a function of time only,
makes it possible to formally derive an effective diffusion coefficient
for RR, $D_{jj}(a,j),$ which is proportional to the power of the
noise at $\nu_{GR}(a,j)$, and is therefore strongly suppressed below
the AI locus. This then makes it possible to directly evolve the stellar
distribution function by the Fokker-Planck (FP) equation. The great
practical advantage of an effective RR diffusion formulation is that
it is then possible to model the dynamics of the loss-cone in the
realistic $\Ns\to\infty$ limit by an MC procedure that solves the
FP equation statistically, and where it is easy to include also NR,
the secular precessions due to GR and the enclosed stellar mass, and
dissipation by GW (and by tidal heating, when relevant). The MC results
are able to reproduce the SB phenomenology observed in the $N$-body
simulations, as well as the plunge and inspiral rates \cite{bar+16}.
Fig. (\ref{f:AISB}) shows how the AI/SB phenomenon is reproduced
by effective RR diffusion, and also how on long timescales the NR
dominates the dynamics, and the system asymptotes to the maximal entropy
solution, as it must irrespective of the randomization mechanism\footnote{The fact the the system is not closed, and there is a small stellar
flux into the less cone, makes possible small deviations from the
maximal entropy solution.}.

\subsection{How fine-tuned is the fortunate coincidence? }

\label{ss:FCfinetune}

\begin{figure}
\begin{centering}
\includegraphics[width=0.75\columnwidth]{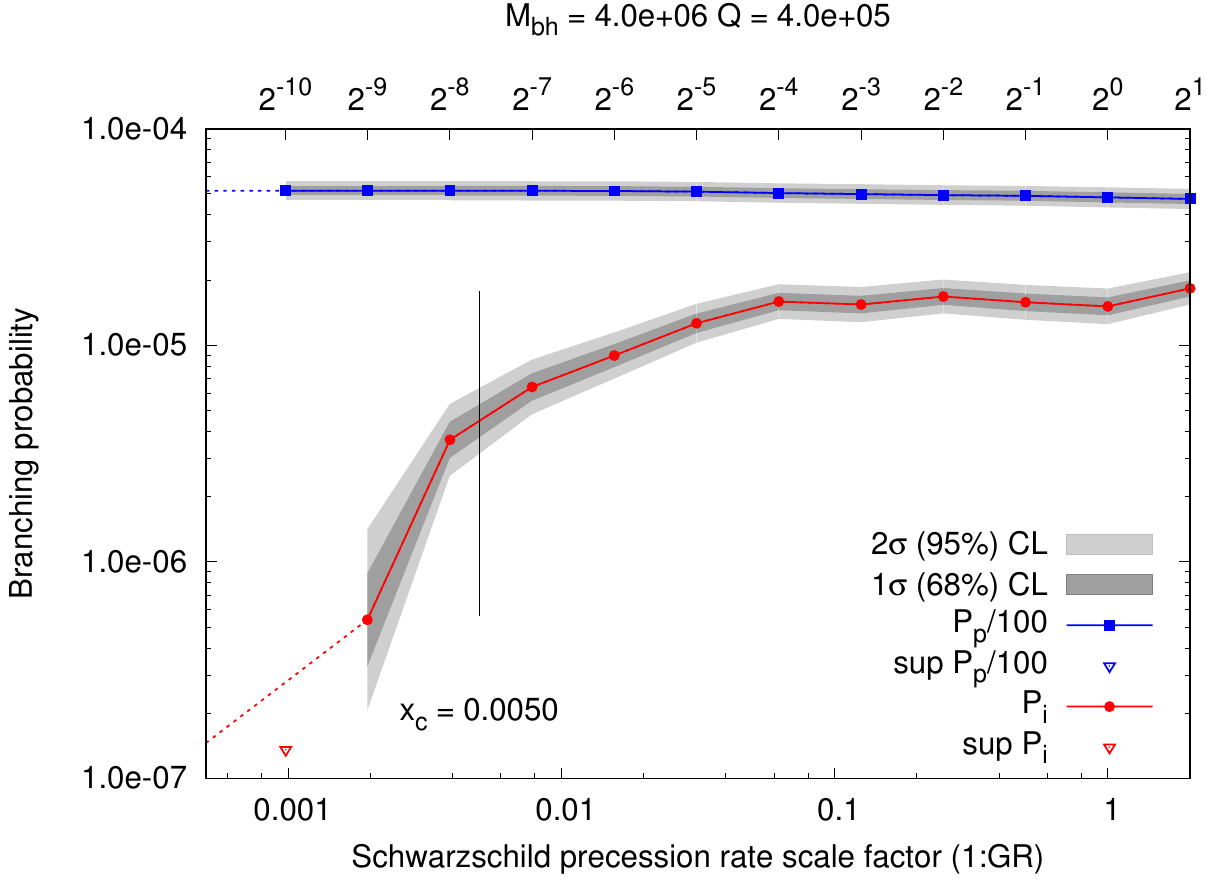}
\par\end{centering}
\caption{\label{f:Pbranch}The run of the plunge and inspiral branching probabilities
with the precession frequency scaling factor $x$ ($x=1$ corresponds
to GR) for a simplified Milky Way model with $\protect\Mbh=4\times10^{6}\,\protect\Mo$
and $\protect\Ms=10\,\protect\Mo$. The blue line is the plunge probability
(divided by $100$ to allow a more compact plot). A dotted blue line
connects it to $P_{p}(x=0)$. The red line is the inspiral (EMRI)
probability. A dotted red line connects it to the result $P_{i}(x=0)=0$.
The gray areas around the lines are the $1\sigma$ and $2$$\sigma$
regions, respectively. The vertical line is the theoretically derived
critical value for $x$ (see text).}
\end{figure}

A simple estimate for the degree of fine-tuning needed for the fortunate
coincidence can be derived by artificially varying the Schwarzschild
precession frequency by a constant factor, $\nu_{x}=x\nu_{GR}$ ($x\ge0$),
and testing how this affects the branching ratio between plunges and
inspirals. In terms of the phase space shown in Fig. (\ref{f:EJ}),
a scaling factor $x<1$ shifts the AI line parallel to itself downward,
and extends the region where RR is effective. A simple criterion for
predicting the effect of the scale factor is the position of the intersection
point of the AI locus and the LSO line relative to the the critical
sma for EMRIs, $a_{GW}$ (Figure \ref{f:EJ}). EMRIs are no longer
possible when RR is effective at all values of $j_{lc}\le j\le1$
down to the critical sma for EMRIs, $a_{GW}$, since then all stars
are swept into the MBH on plunging trajectories before they can diffuse
below $a_{GW}$. This occurs when the AI locus $j_{0}(a;x)=\sqrt{T_{c}(a)\nu_{x}(a,j=1)/2\pi}$,,
crosses the plunge loss cone $j_{\bullet}(a)$ below $a_{GW}\simeq0.03(\log Q)^{-4/5}r_{h}$
\cite{bar+16}, where a relaxed cusp with $\alpha=7/4$ is assumed,
and where the numeric prefactor depends on the approximation of the
GW dissipation. The scaling of $r_{h}$ with $Q$ (assuming an $\Mbh/\sigma$
relation $\Mbh\propto\sigma^{4}$ and fixed $\Ms$) is $r_{h}\propto Q^{1/2}$,
and so $a_{GW}\propto(\log Q)^{-4/5}Q^{1/2}$. The condition $j_{0}(a_{GW};x_{c})/j_{lc}(a_{GW})=1$
then defines a critical value $x_{c}$ such that for $x<x_{c}$, RR
remains unquenched everywhere above $a_{GW}$. Fig. (\ref{f:Pbranch})
shows the run of the branching ratio with $x$ for the Milky Way model
of Fig. (\ref{f:EJ}). The ratio changes in favor of plunges (cf.
Fig. \ref{f:FC}) for $x<x_{c}\sim{\cal O}(5\times10^{-3})$. This
suggests that EMRIs are robust in the context of Einstein's GR.

\section{Applications of the $\eta$-formalism}

\label{s:applications}

\subsection{Cosmic loss-rates}

\begin{figure}
\noindent \begin{centering}
\includegraphics[width=0.75\columnwidth]{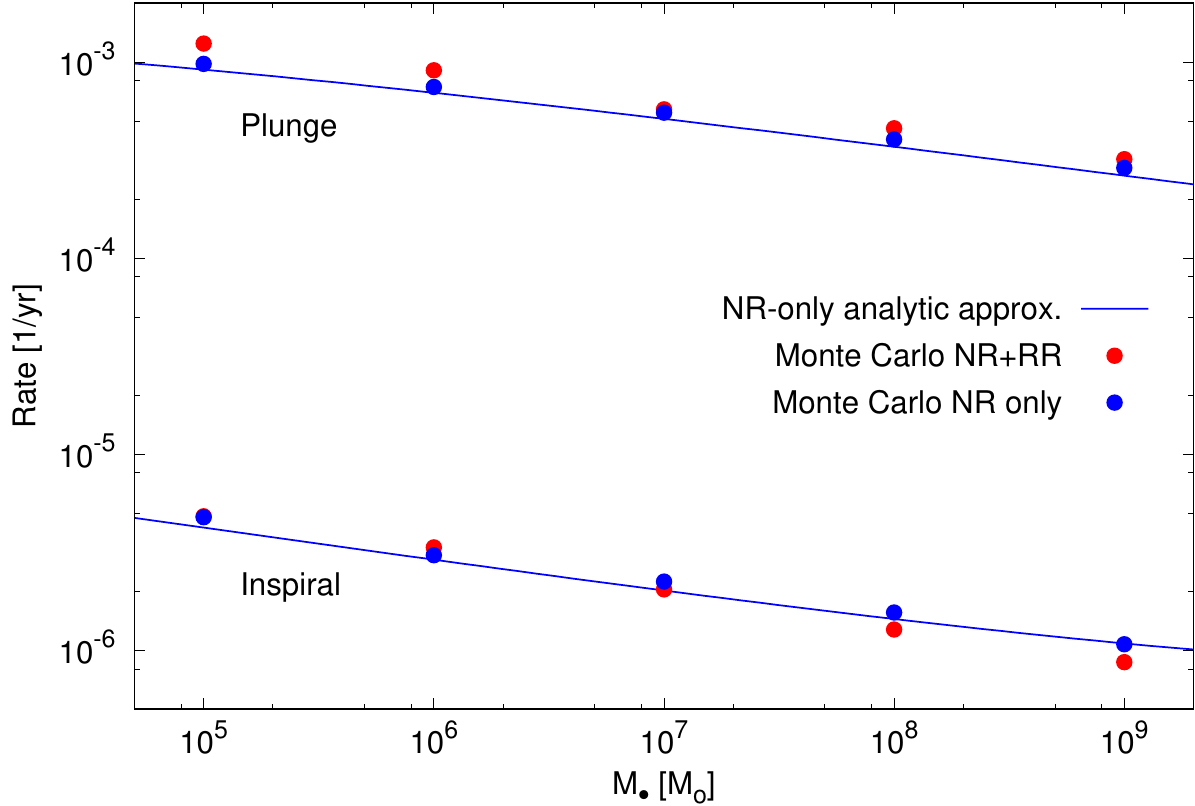}\caption{\label{f:rates}The plunge and inspiral rates, as function of the
MBH mass $\protect\Mbh$, as derived from Monte Carlo simulations
of the loss-cone in a sequence of simplified galactic nuclei models
(adapted from \cite{bar+16}). The presence of RR makes only a small
difference to the rates, since the loss lines (cf Figure \ref{f:EJ})
are well below the AI line (Eq. \ref{e:jAI}). The Inspiral rates
are $<0.01$ smaller than the plunge rate (Sec. \ref{ss:losscone}),
and both fall roughly as $\protect\Mbh^{-1/4}$ (for an $\protect\Mbh/\sigma$
relation with $\beta=4$).}
\par\end{centering}
\end{figure}

MC simulations such as the one presented in Fig. (\ref{f:FC}) for
$\Mbh=4\times10^{6}\,\Mo$ can be carried out for any value of $\Mbh$
once the properties of the galactic nucleus are related to the MBH
mass via the $\Mbh/\sigma$ relation. Fig. (\ref{f:rates}) Shows
the run of the steady state plunge and EMRI inspiral rates as function
of the MBH mass \cite{bar+16}, and demonstrates the fact that RR
plays only a small role in the rates, due to the very effective AI
suppression of RR well away from the plunge and inspiral loss-lines
(Fig. \ref{f:EJ}).

\subsection{Loss-cone dynamics around the MBH of the Milky Way}

\label{ss:Sstars}

\begin{figure}
\noindent \begin{centering}
\begin{tabular}{cc}
\includegraphics[width=0.5\columnwidth]{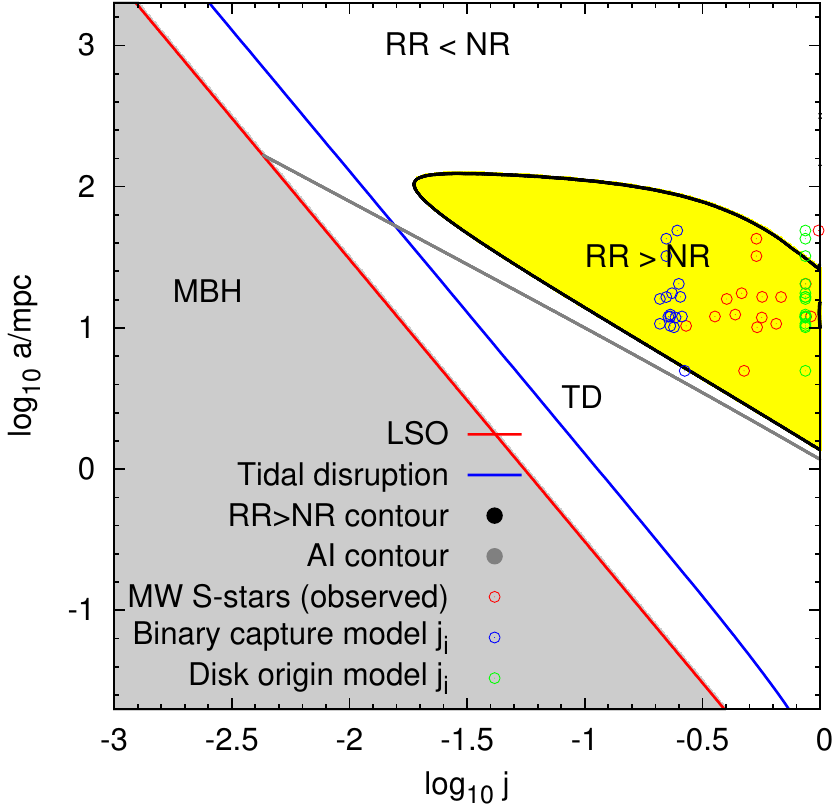} & \includegraphics[width=0.5\columnwidth]{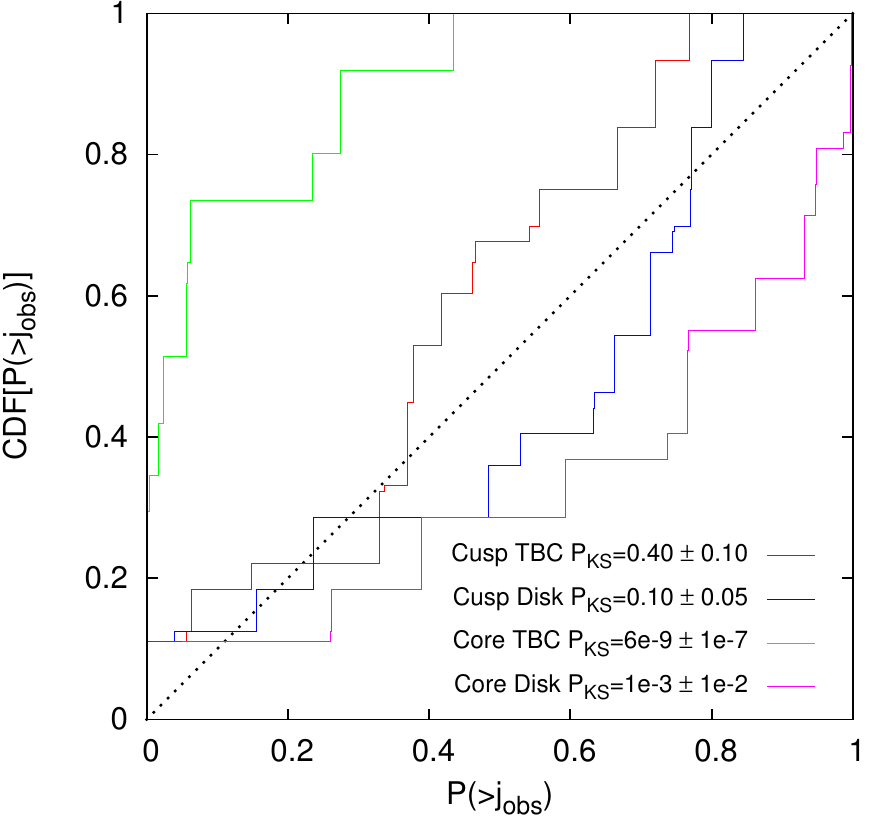}\tabularnewline
\end{tabular}
\par\end{centering}
\caption{\label{f:Sstars}A joint test for the existence of a high density
stellar cusp around $\protect\SgrA$ and for the origin of the S-stars
that are observed to orbit the MBH in the central $\sim0.04$ pc.
Left: The S-stars \cite{gil+09} (red circles) lie in the phase space
region where RR dominates the dynamics (the tidal disruption line
for a typical S-star is shown in blue). If they originated from tidal
binary captures (TBC, the Hills mechanism \cite{hil88}), they start
their existence near the MBH on highly eccentric orbits (blue circles),
at an arbitrary phase of their lifespan. However, if the S-stars originated
from the young ($\sim6\times10^{6}\,\mathrm{yr}$ \cite{bar+09})
stellar disk around observed around $\protect\SgrA$, they are only
as old as the disk, and are expected to originate on nearly circular
orbits (green circles). Right: A comparison of the final cumulative
$j$-distribution of the S-stars for the two formation scenario, as
derived from MC simulations for two possible stellar distributions
around the MBH: a dense relaxed cusp, and an out-of-equilibrium lower
density stellar core. The best fit scenario (highest K-S probability
with distribution closest to a straight line) is the tidal binary
capture in a high density stellar cusp (Sabsovich, Alexander \& Bar-Or,
in prep.).}

\end{figure}

One intriguing result that follows from the formal treatment of effective
RR diffusion, and of the identification of the phase space region
where RR dominates the dynamics, is that almost all the observed S-stars
around $\SgrA$ \cite{gil+09} (main sequence B stars, with masses
$\sim5-20\,\Mo$ and lifespans $\sim10^{7}-10^{8}$ yr) are found
inside the strong RR region (Fig. \ref{f:Sstars}). This suggests
that whatever the unknown formation (or capture, or migration) process
that is responsible for the puzzling presence of young massive stars
so deep in the potential of a MBH, it is likely that they have undergone
strong post-formation $j$-evolution. Fig. (\ref{f:Sstars}) shows
a preliminary attempt to discriminate between the two leading models
for the origin of the S-stars: Capture by a 3-body tidal interaction
of the MBH with an incoming massive stellar binary \cite{gou+03,per+07}
(tidal binary capture, or the Hills mechanism \cite{hil88}), or migration
from the observed stellar disk around $\SgrA$ \cite{lev07}. In addition,
such models also probe the unobserved population of faint stars and
remnants around the MBH, which are necessary for generating the RR
torques, but whose presence is observationally controversial \cite{do+09,buc+09,bar+10}
(but see recent detection of a cusp of faint low mass stars in the
Galactic Center \cite{gal+17,sch+17}). The MC experiments indicate
that the likeliest scenario is that the S-stars were captured in tidal
binary separation events, and their orbits subsequently evolved in
the presence of high density ``dark'' stellar cusp. 

\section{Summary}

\label{s:summary}

The complex dynamics that ultimately lead stars and compact remnants
to fall into the MBH can be modeled by MC simulations that introduce
RR by effective diffusion coefficients that are derived from the $\boldsymbol{\eta}$-formalism.
Phase space is clearly separated into a restricted region where RR
dominates the dynamics, and a region where AI strongly suppresses
RR due to fast precession. Elsewhere, RR exists, but NR is faster.
Importantly, both the the loss-lines for plunges (LSO) and for GW
EMRIs lie below the AI locus, and so RR has only a small effect on
the loss-rates. This situation appears to be a fortunate coincidence
for the prospects of detecting EMRIs. It is a result of a three-way
competition between RR torques that tend to push stars to plunge orbits,
GR Schwarzschild precession, which suppresses RR, and GR GW dissipation,
which when fast enough, can be completed before the stellar background
interferes. This coincidence appears robust in the context of Einstein's
GR. It is however still unclear how general it is in the wider context
of theories of strong gravity.

The dynamical processes that play part in EMRI dynamics can be probed
by the puzzling S-stars observed near the MBH of the Galactic Center,
which are at the RR-dominated region of the phase space around $\SgrA$.

\ack

This work was supported by the I-CORE Program of the Planning and
Budgeting Committee and The Israel Science Foundation (grant No 1829/12). 

\section*{References}

\bibliographystyle{iopart-num}
%\bibliography{procs}
\providecommand{\newblock}{}

\end{document}